\documentstyle[epsfig]{article}
\def\be{\begin{equation}}
\def\ee{\end{equation}}
\def\bea{\begin{eqnarray}}
\def\eea{\end{eqnarray}}

\def\fs{8.5}

\begin{document}

\title{Measurement of mechanical vibrations excited in aluminium 
resonators by 0.6~GeV electrons.
}
\author{G.D.~van Albada$^{a}$,
E.~Coccia$^{b}$, V.~Fafone$^{c}$,
H.~van der Graaf$^{d}$, \\ G.~Heijboer$^{d}$,
J.W.~van Holten$^{d}$, W.J.~Kasdorp$^{d}$,
J.B.~van der Laan$^{d}$,\\ L.~Lapik\'{a}s$^{d}$, G.~Mazzitelli$^{c}$,
G.J.L.~Nooren$^{d}$, C.W.J.~Noteboom$^{d}$,\\ J.E.J.~Oberski$^{d}$,
G.~Pallottino$^{e}$,
H.Z.~Peek$^{d}$, F.~Ronga$^{c}$,\\ A.~Schimmel$^{d}$, T.G.B.W.~Sluijk$^{d}$,
P.~Steman$^{d}$, J.~Venema$^{d}$,
P.K.A.~de~Witt~Huberts$^{d}$.
\vspace{1cm}
\\{\small a) Dept. of Computer Science, Univ. van Amsterdam;, The Netherlands}
\\{\small b) Dept. of Physics, Univ. of Rome, "Tor Vergata", and INFN, Italy;}
\\{\small c) Lab. Nazionale di Frascati, INFN, Italy;}
\\{\small d) NIKHEF, P.O.B. 41882, 1009 DB Amsterdam, The Netherlands;}
\\{\small e) Dept. of Physics, Univ. of Rome "La Sapienza", and INFN, Italy}
\\{email: J.Oberski@nikhef.nl}
}
\maketitle
{\Large
{\begin{center} 
 {NIKHEF 99-036 
  \em To be published by Review of Scientific Instruments, May 2000}
 \newline
 \newline Abstract\end{center}

We present measurements of 
mechanical vibrations induced by 0.6 GeV electrons impinging on 
cylindrical and spherical aluminium resonators.
To monitor the amplitude of the resonator's vibrational modes
we used piezoelectric ceramic sensors,
calibrated by standard accelerometers. 
Calculations using the thermo-acoustic conversion model,
agree well with the experimental data,
as demonstrated 
by the specific variation of the excitation strengths with the
absorbed energy, and with the traversing particles' track positions.
For the first longitudinal mode of the cylindrical resonator
we measured a conversion
factor of 7.4~$\pm$~1.4~nm/J, confirming the model value of 10~nm/J.
Also, for the spherical resonator,
we found the model values for the $L$=2 and $L$=1 mode amplitudes
to be consistent with our measurement.
We thus have confirmed the applicability of the model,
and we note that calculations based on the model have shown
that next generation 
resonant mass gravitational wave detectors can only be expected to
reach their intended ultra high sensitivity if they will be shielded
by an appreciable amount of rock, where a
veto detector can reduce the background of remaining impinging cosmic rays
effectively.
}

{\small 04.80.Nn 07.07.Df 07.64.+z 07.77.Ka 29.40.Gx 29.40.Wk
43.20.Ks 43.35.Ud 95.55.Ym 96.40.Vw 96.40.z}

\pagebreak
\section{Introduction}
\label{sec:introduction}
\Large
A key issue for a Resonant Mass
Gravitational Wave
Detector ~\cite{grail} of improved sensitivity 
with respect
to the existing detectors,
is the background due to impinging cosmic ray particles
\cite{ricci,mazzi-1}.
The energy deposited in the detector's mass  
along a particle's track may excite
the very vibrational modes that are to signal
the passing of a gravitational wave.
Computer simulations of such effects
are based on the 
thermo-acoustic conversion model
and earlier measurements of resonant effects in
Beron et al.~\cite{beron} and Grassi Strini et al.~\cite{grassi}. 
According to the model, the energy deposited 
by a traversing particle
heats the material locally around the particle track,
which leads to mechanical tension and thereby
excites acoustic vibrational modes \cite{askar}.
At a strain sensitivity of the order 
of $10^{-21}$ envisaged for a next generation
gravitational-wave detector, computer 
simulations ~\cite{mazzi-1,obers} show that operation of the instrument
at the surface of the earth would
be prohibited by the effect of the cosmic ray background.
Since the applicability of the thermo-acoustic conversion model 
would thus yield an important constraint on
the operating conditions of
resonant mass gravitational wave detectors,
Grassi-Strini, Strini and Tagliaferri \cite{grassi} measured
the mechanical vibrations in a bar resonator
bombarded by 0.02~GeV protons and 5*10$^{-4}$~GeV electrons.
We extended that experiment by measuring the excitation patterns in
more detail for a bar and a sphere excited by 0.6~GeV electrons.
Even though we cannot think
of a reason why the model, if applicable to the bar, would not hold
for a sphere, we did turn to measuring with a sphere also. 
\newline
We exposed \cite{ref:amaldi} two aluminium 50ST alloy cylindrical bars and 
an aluminium alloy sphere, each
equipped with piezoelectric ceramic sensors, 
to a beam of $\approx 0.6$~GeV
electrons used in single bunch mode with a pulse width of
up to $\approx 2 \mu $s, and adjustable intensity of
$10^9$ to $10^{10}$ electrons.
We recorded the  
signals from the piezo sensors,
and Fourier analysed their time series. 
Before and after the beam run we calibrated the sensor response of
one of the bars 
for its first longitudinal vibrational mode at
$\approx $~13~kHz 
to calibrated accelerometers.

\section{Experiment setup and method}
\label{sec:setup}
In the experiment we used three different setups in various runs, as 
summarised in table~I: two bars and a sphere.
With the un-calibrated bar BU we explored the feasibility of the
measurement.
Also, bar BU proved useful to indirectly determine
the relative excitation amplitudes of
higher longitudinal vibrational modes, see sec. \ref{sub:bar}.
With bar BC calibrated at its
first longitudinal vibrational mode,
we measured directly its excitation amplitude in the beam.
Finally, with the sphere we further explored the applicability of the model.
\newline

\begin{minipage}{15cm}
\small
\vspace{1 cm}
\begin{center}
{\Large Table I. Characteristics of our setup.}
\begin{tabular}{|l|c|c|c|}
\hline
Setup code name: & BC & BU & SU \\
\hline 
Resonator type: & bar & bar & sphere \\
Diameter: & 0.035 m & 0.035 m & 0.150 m \\ 
Length: & 0.2 m & 0.2 m & -   \\
Suspension: & plastic string & plastic string & brass rod 0.15~m*0.002~m \\
Piezo sensors:  & 1 & 2 & 2 \\
Piezo hammer: &  0  &  1 & 1 \\ 
Capacitor driver & 1 & 0 & 0 \\
Direct calibration & yes & no & no \\
Beam energy & 0.76 GeV & 0.62 GeV & 0.35 GeV \\ 
Beam peak current & 3 mA & 18 mA & 19 mA \\
Electrons per burst & $\approx 10^9$ & $\approx 5\cdot 10^{10}$ & 
    $\approx 5\cdot 10^{10}$ \\
Mean absorbed energy per electron  & 0.02 GeV & 0.02 GeV & 0.1 GeV \\
Typical absorbed energy per burst & 0.01 J & 0.6 J & 3.0 J \\ 
\hline 
\end{tabular}
\end{center}
\end{minipage}

\Large
\subsection{Electron beam}
\label{sub:beams}
We used the Amsterdam linear electron accelerator 
MEA~\cite{strooi,meaweg}
delivering
an electron beam with a
pulse-width of up to 2~$\mu $s in its hand-triggered, single bunch mode. 
The amount of charge per beam pulse was varied,
recorded by a calibrated digital oscilloscope,
photographed and analysed off line to determine the number of impinging
electrons per burst.
\subsection{Suspension and positioning}
\label{sub:suspension}
In both setups BC and BU, see fig.~1, 
the cylindrical
aluminium bar was
horizontally suspended in the middle, as indicated in the figure,
with a plastic string.
\begin{figure}[h]
\begin{minipage}[l]{17 cm}
\psfig{figure=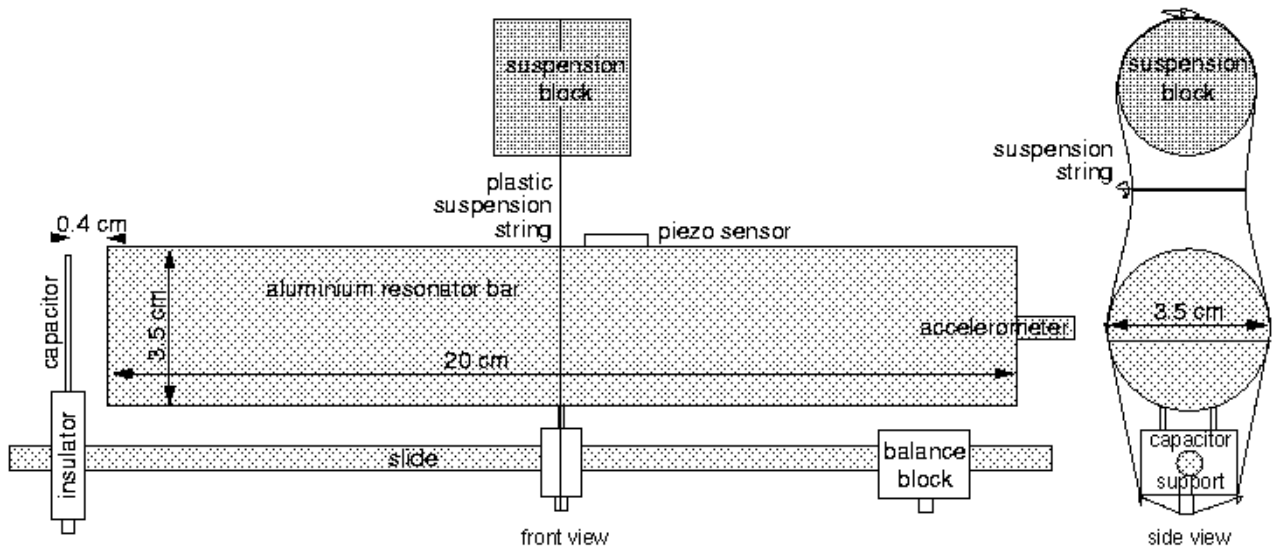,width=17 cm}
\label{fig:barBC}
\caption{\vspace{1.0cm}
Calibrated bar setup BC. The beam traverses the bar's front side
perpendicular to the drawing plane.
}
\end{minipage}
\end{figure}
The bar's cylinder axis
was positioned at 90$^0$ to the beam direction. 
The bar's suspension string was connected to a 
horizontally movable gliding construction,
enabling us to handle the resonator 
by remote control, and let the impinging electron beam hit it
at different horizontal positions. 
The aluminium
sphere SU, see fig.~2,
was suspended from its centre by a 
brass rod.
Either the bar's gliding construction, or the sphere's suspension bar,
was attached to an aluminium tripod mounted
inside a vacuum chamber~\cite{strooi}, which was
evacuated to about $10^{-5}~$mbar.
By remote control, we rotated the tripod and moved it
vertically to either let the beam
pass the resonator
completely, or let it traverse the resonator. 
We let the beam traverse the sphere at different heights and 
different incident angles with respect to the piezo sensors positions on the
sphere. 
\begin{figure}[h]
\vspace{-0.4cm}
\begin{center}
\begin{minipage}[c]{\fs cm}
\psfig{figure=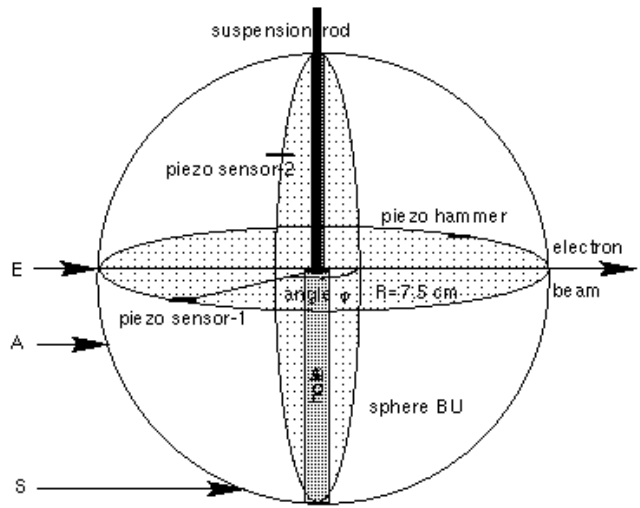,width=\fs cm}
\label{fig:sphereSU}
\caption{\vspace{1.0cm}
The spherical resonator setup, SU.
}
\end{minipage}
\end{center}
\end{figure}
We mark the beam heights as E (Equator) and A (Africa) at 0.022~m
below the equator.
The E beam passed horizontally through the sphere's origin, remaining in the
same vertical plane for the A beam.
\subsection{Sensors and signal processing}
\label{sub:sensors}
In setup BC
we used a single piezo sensor of $\approx 15*3*1~$mm$^3$
and glued it over it full length 
at 0.01~m off the centre
on the top of the bar.
Bar BC was equipped with a capacitor plate of 0.03~m diameter
at a distance of $\approx 0.004$~m from one of its end faces. 
In setup BU
one piezo sensor of $\approx~3*6*0.3~$mm$^3$
was fixed on one end face of the bar.
\newline
A similar sensor of about the same dimensions
was fixed in the same manner, oriented parallel to the cylinder's
long axis at a position 35~mm away from the end-face,
In the third setup, SU, see fig.~2,
two piezo sensors of $\approx~3*6*0.3~$mm$^3$
were glued to the sphere's surface.
One was situated at the equator, with respect to the vertical
rotation axis,
the other one at a relative displacement of
$45^0$ west longitude, and at $45^0$ north latitude. 
\newline
For the setup in use, each sensor was connected to
a charge amplifier of $\approx 2*10^{10}~$V/C gain.
The signals were sent through a Krohn-Hite 3202R low-pass 100~kHz
pre-filter, to a R9211C Advantest spectrum analyser with internal
2~MHz pre-sampling and 125~kHz digital low-pass filtering.
The oscillation signals were recorded for 64~ms periods at a
4~$\mu$s sample rate.
The beam pulse could be used as a delayed trigger to the Advantest.
Using the memory option of
the Advantest, 
the piezo signals were recorded from ~0.3~ms onward 
{\em before} the arrival of the trigger. 
The data were stored on disk and were Fourier analysed off line.
\newline
\subsection{Checks and stability}
\label{sub:stability}
The data were taken at an ambient temperature of $\approx 23~^0$C.
By exciting the resonator with the piezo-hammer we checked roughly its overall
performance.
As to be discussed in section \ref{sec:calibration},
setup BC was calibrated before and after the beam run. 
The instrument's stability was checked several times 
during the run by an electric driving signal on its capacitor endplate.

\section{Calibration of bar BC's piezo ceramic sensor}
\label{sec:calibration}
A standard accelerometer mounted on the bar damped the vibrations
too strongly to confidently measure their excitations in the electron beam.
Therefore 
the response of the piezoelectric ceramic together with its amplifier
was first calibrated against
two 2.4~gramme Bruel\&Kjaer 4375
accelerometers 
glued, one at a time, to bar BC's end face and
connected to a 2635 charge amplifier. 
The resonator was excited through
air by a nearby positioned loud-speaker 
driven from the Advantest digitally tunable sine-wave generator.
The output signals from both the piezoelectric ceramic
amplifier and the accelerometer amplifier were fed into the Advantest.
Stored time series were read out by an
Apple Mac 8100 AV, running Lab-View for on-line
Fourier analysis, peak selection, amplitude and decay time determination.
\newline
We took nine calibration runs
varying the charge amplifier's sensitivity setting, and
dismounting and remounting either of the two accelerometers to the bar.
For the lowest longitudinal vibrational mode we calculated
the ratio of the Fourier 
peak signal amplitudes, $R$, from the piezoelectric ceramic
and accelerometer.
\newline
With the calibrated bar BC positioned in the electron beam line
we checked the stability of the piezoelectric ceramic's response intermittently
with the beam runs by exciting the bar through its capacitor
plate at one end face, electrically driving it at and around
half the bar's resonance frequency. We found the response to remain stable
within a few percent.
\newline
After the beam runs we took additional calibration values in air with a newly
acquired
Bruel\&Kjaer 0.5~g 4374S subminiature accelerometer
and a Nexus 2692 AOS4 charge
amplifier.
\begin{figure}[h]
\vspace{-0.5cm}
\begin{minipage}[l]{\fs cm}
\psfig{figure=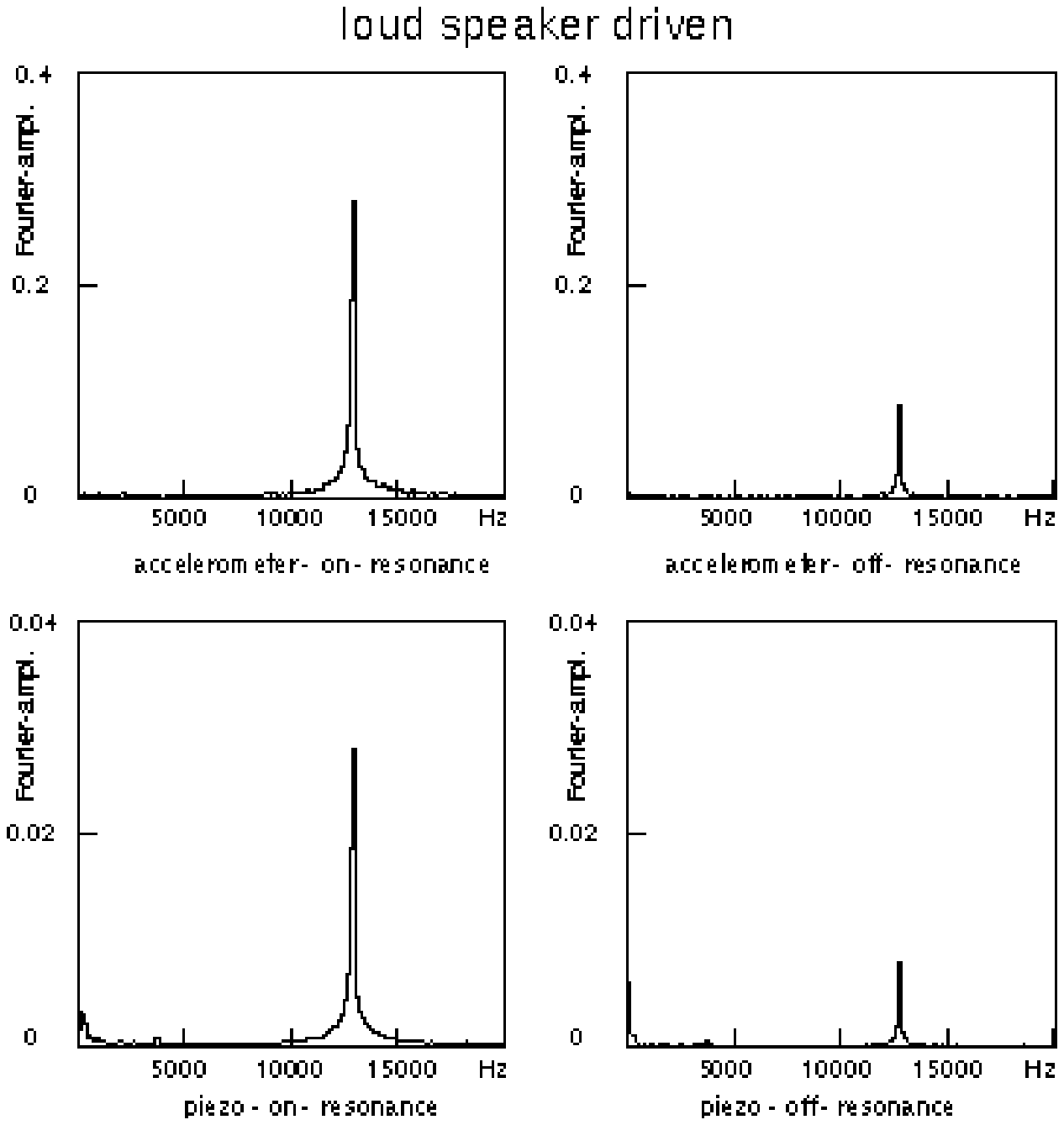,width=\fs cm}
\label{fig:calBC-a}
\caption{
Response at the lowest longitudinal acoustic frequency 
of bar BC's piezo (lower) and accelerometer (upper)
Fourier amplitudes by constant amplitude loud-speaker driving.
Left: on-resonance, $f$=12950 Hz.
Right: slightly off-resonance, $f$=12850 Hz.
}
\end{minipage}
\begin{minipage}[r]{\fs cm}
\vspace{0.6cm}
\psfig{figure=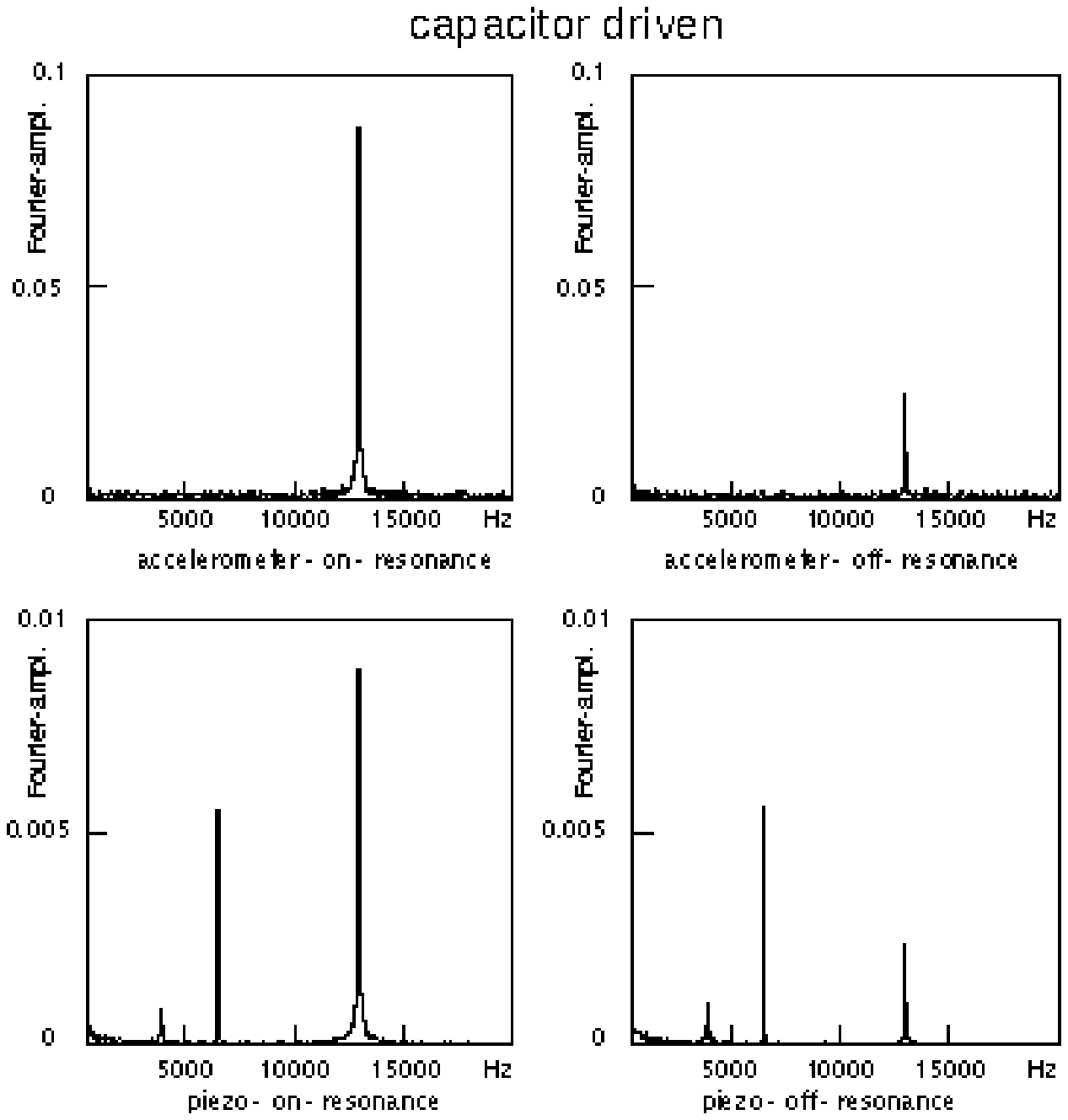,width=\fs cm}
\label{fig:calBC-b}
\caption{
Response of bar BC's piezo (lower) and accelerometer (upper) Fourier
amplitudes by electrostatic capacitor plate driving.
Left: on-resonance, $f$=6481.7 Hz. Right: slightly off-resonance, 
$f$=6480.0 Hz. 
The piezo peak of constant amplitude
at 6.5 kHz arising from crosstalk is absent in the
accelerometer, while the acoustic resonance is clearly seen at 13 kHz in both.
}
\end{minipage}
\end{figure}
In fig.~3, typical frequency responses
are shown when driving the bar by a loud-speaker signal.
The upper part gives the Fourier peak amplitude of
the bar's 13 kHz resonance as measured with the accelerometer.
The lower part gives the corresponding amplitude for the signal from the
piezoelectric ceramic. 
The right hand side
of the picture shows the amplitudes to be smaller, as expected
when driving  the bar slightly off resonance.
We calculate the decay time, $\tau$, of the k-th mode amplitude
$A_k(t)=A_k(0)\cdot e^{-t/\tau}$
to be $\tau=0.4$~s
for this setup, equipped with
the relatively light accelerometer. 
\newline
Figure~4 
shows the corresponding two signals 
when driving the bar by the capacitor plate at 6.5~kHz, that is at
half the bar's resonance frequency. Here,
the direct electric response of the piezoelectric ceramic's signal
to the driving sine-wave is present, clearly
without a mechanic signal, as would have shown up in the accelerometer. 
The direct signal at 6.5~kHz remains constant. On the other hand,
the bar's mechanical
signals on and off its resonance frequency around 13~kHz
show the expected amplitude change again, thereby demonstrating that around
the bar's resonance, the
piezoelectric ceramic does only
respond to the mechanical signal, not to the electric driving signal.
See also the caption of fig.~4.

We calculated the average value of
$R_0=V_{piezo}^{Fourier}/V_{accel.}^{Fourier}$ 
and the error over all 29 measurements,
finding for the calibration factor at f=13~kHz, 
\be
\beta= R_0S(2\pi f)^2=  (2.2~\pm~0.3)~\mbox{V/nm}
\label{eq:ijk}
\ee
where $S=0.1~$V/ms$^{-2}$ is the amplifier setting of the accelerometer.

\section{Beam experiments}
\label{sec:results}
Sensor signals way above the noise level were
observed for every beam pulse hitting the sphere
or the bar.
We ascertained that: a) the signals arose from mechanical
vibrations in the resonator, and b) they
were directly initiated by the effect of
the beam on the resonator, and not arising from an indirect 
effect of the beam on the piezo
sensors. Our assertion is based on a combination
of test results observed 
for both the bars and the sphere, as now to be discussed.

\begin{figure}[h]
\begin{minipage}[l]{\fs cm}
\psfig{figure=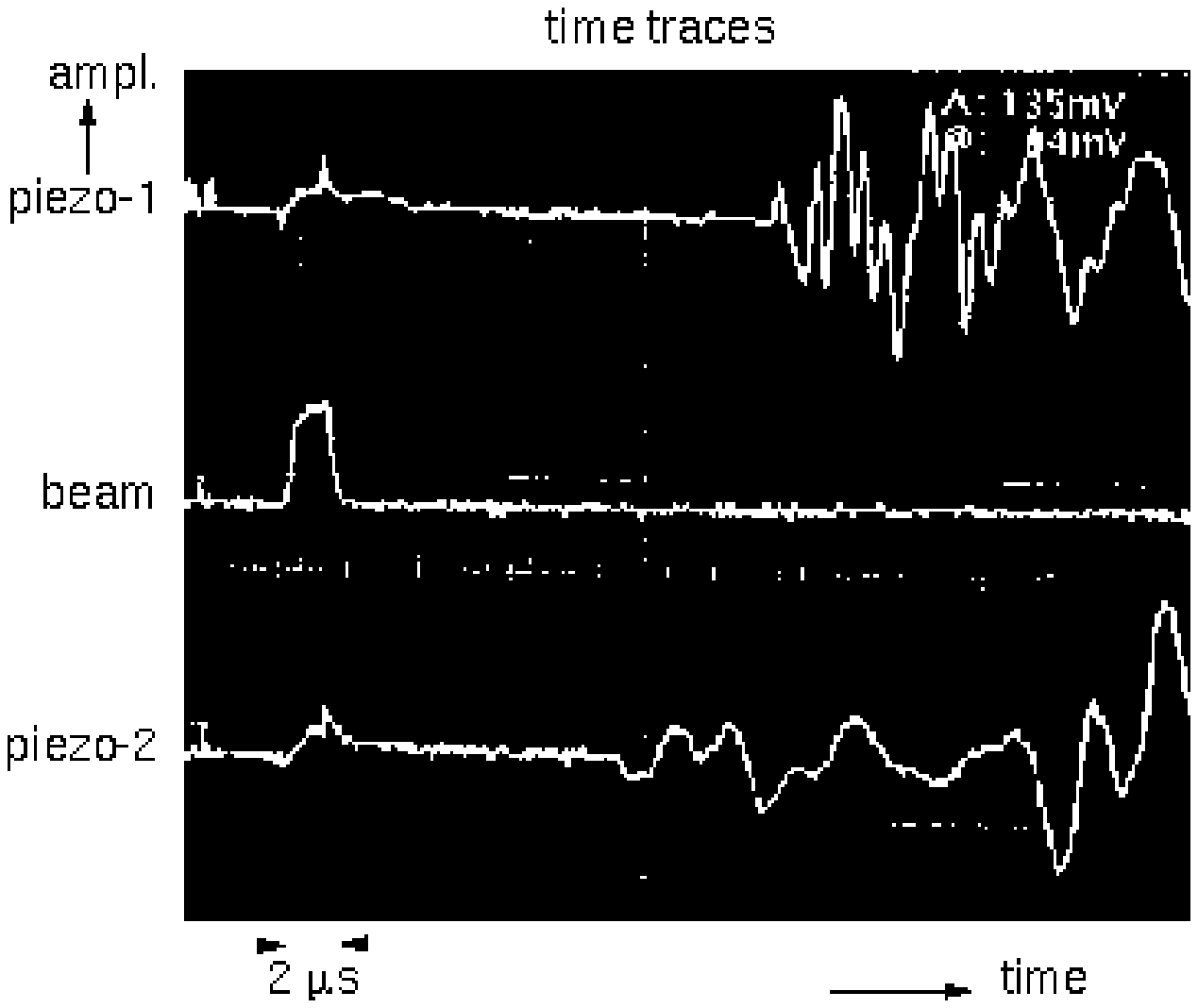,width=\fs cm}
\label{fig:foto-1}
\caption{
The $2~\mu$s electron bunch (middle trace) and
the piezo sensors signals on the sphere (upper and lower traces).
}
\end{minipage}
\begin{minipage}[r]{\fs cm}
\psfig{figure=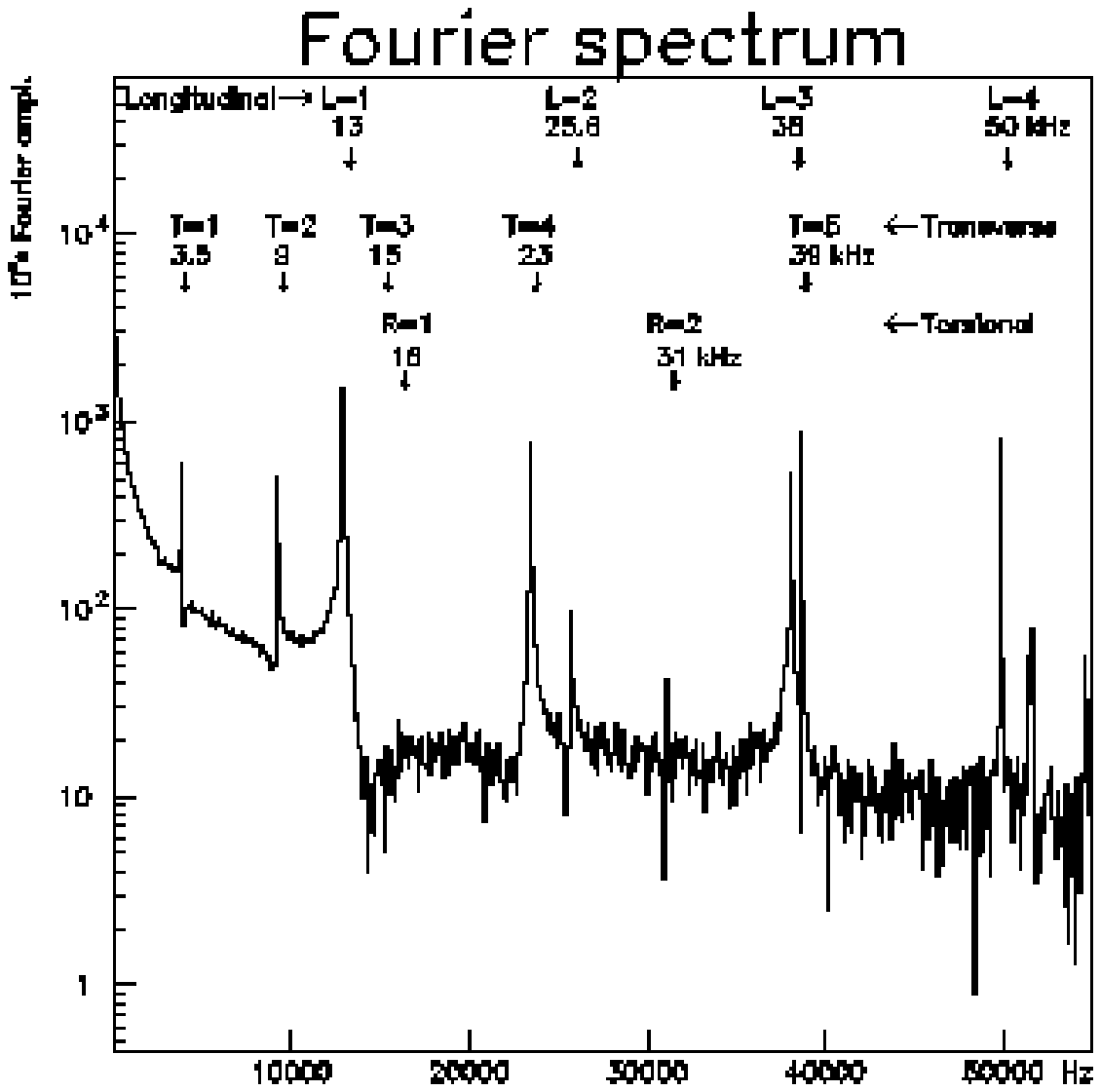,width=\fs cm}
\label{fig:barFourier}
\vspace{-0.6cm}
\caption{
A typical Fourier spectrum of bar BC as excited by the electron beam.
Data were analysed over 0.016~s duration,
from 0.008~s onwards after the beam passed.
Identified vibrational modes are indicated. 
}
\end{minipage}
\end{figure}
First, when the beam passed underneath the resonator
without hitting it, we observed no
sensor signal above the noise.
Second, as shown in fig.~5, the sensors'
delayed responses after the impact of the beam agreed with the sound velocity.
Here the beam was hitting the sphere at a position 
5~mm above the sphere's south pole. 
The middle trace shows the beam pulse of
$\approx 2\mu$s duration. 
The two other traces show
both piezo sensors to respond with
a transient signal right from the start time of the beam's arrival 
and to begin oscillating after some delay, depending on their distance from the
beam. 
The distance of the equatorial
sensor to the beam hitting the sphere at the south pole 
was 0.11~m, corresponding to $\approx 22~\mu$s travel time
for a sound velocity of $\approx5*10^3$~m/s. 
The signal
is indeed seen in the lowest trace starting to oscillate at that
delay time.
The upper trace shows the signal from the second sensor situated on
the northern hemisphere
at 0.14~m from the traversing beam, correspondingly
starting to oscillate with a delay of $\approx 28~\mu$s 
after the impact of the beam.
Third, after dismounting the piezo-hammer from the resonator, 
we observed that the sensor
signals did not change, which showed 
that the activation is not caused by the beam inducing a
triggering of the piezo-hammer.
Fourth, to simulate the electric effect of the beam pulse on the sensors,
we coupled a direct current of 60~mA
and 2.5~$\mu$s duration
from a wave packet generator
to the bar. Apart from the direct response of the piezo-sensor during the input
driving wave, no oscillatory signal was detected above the noise level.
Finally, we measured 
the dependence of the amplitudes in several vibrational modes
on the hit position of the beam,
as will be described in the following sections.
We found the amplitudes to
follow the patterns as calculated with the thermo-acoustic conversion model.
\newline

\noindent
\subsection{Results for the bar}
\label{sub:bar}
In fig.~6 a typical Fourier spectrum of bar BC is shown up
to 55~kHz.
The arrows point to identified vibrational modes \cite{DvA-bar}.
From a fit of $K$ and $f_0$ 
of the longitudinal frequencies $f_L=L\cdot f_0\cdot (1 - L^2 K)$
\cite{freq-long}
of the modes for $L$=1,..,4,  
we find $f_0=12933, K=0.0022$, where $f_0$ is related to the sound 
velocity by $v_s=2l*f_0=5173$~m/s for our bar length of $l$=0.2~m.
For the Poisson-ratio 
$\sigma = 2l\surd K/(\pi r)$, $r$ being the cylinder radius of the bar,
from our fit we get $\sigma =0.338$. The values agree well with the handbook
\cite{metal-hb} quoting $\sigma$=0.33 and $v_s$=5000 m/s for aluminium.
The root mean square error of the fit is 35~Hz, in correspondence with
the 30~Hz frequency resolution used in the Fourier analysis.
Other peaks correspond
to torsional and transverse modes~\cite{DvA-bar,freq-long}.

\begin{figure}[h]
\vspace{-0.4cm}
\begin{minipage}[l]{\fs cm}
\vspace{3.5cm}
\psfig{figure=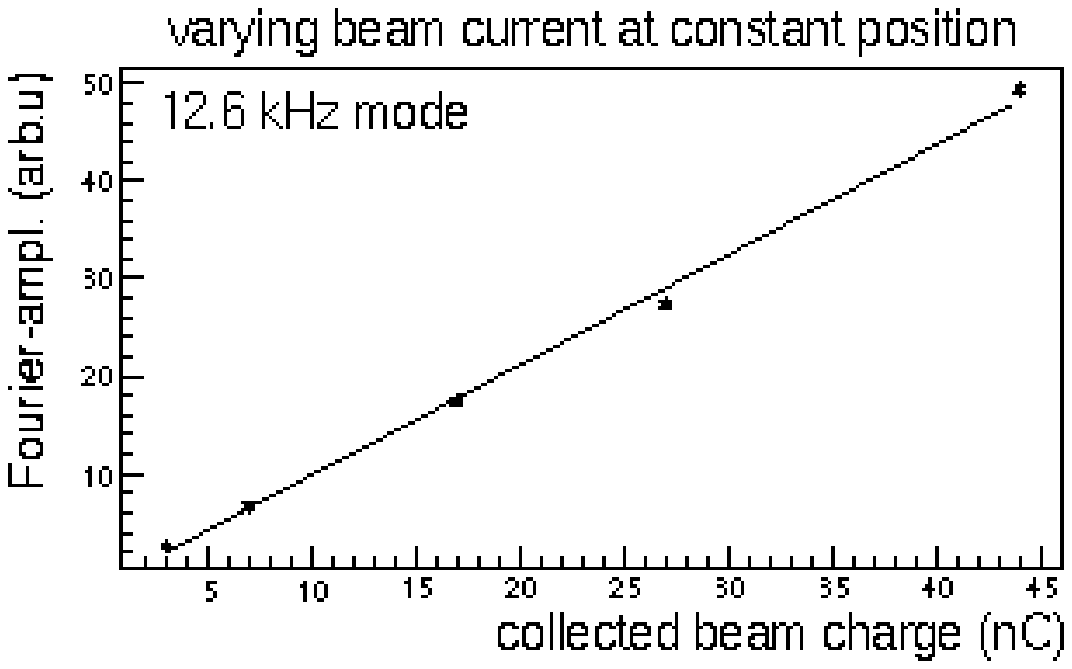,width=\fs cm}
\vspace{-1.0cm}
\caption{
Correlation between the Fourier amplitude of the 12.6~kHz vibrational mode
and the beam charge. Data points (*) and straight line fit.
}
\label{fig:FvsE}
\end{minipage}
\begin{minipage}[r]{\fs cm}
\psfig{figure=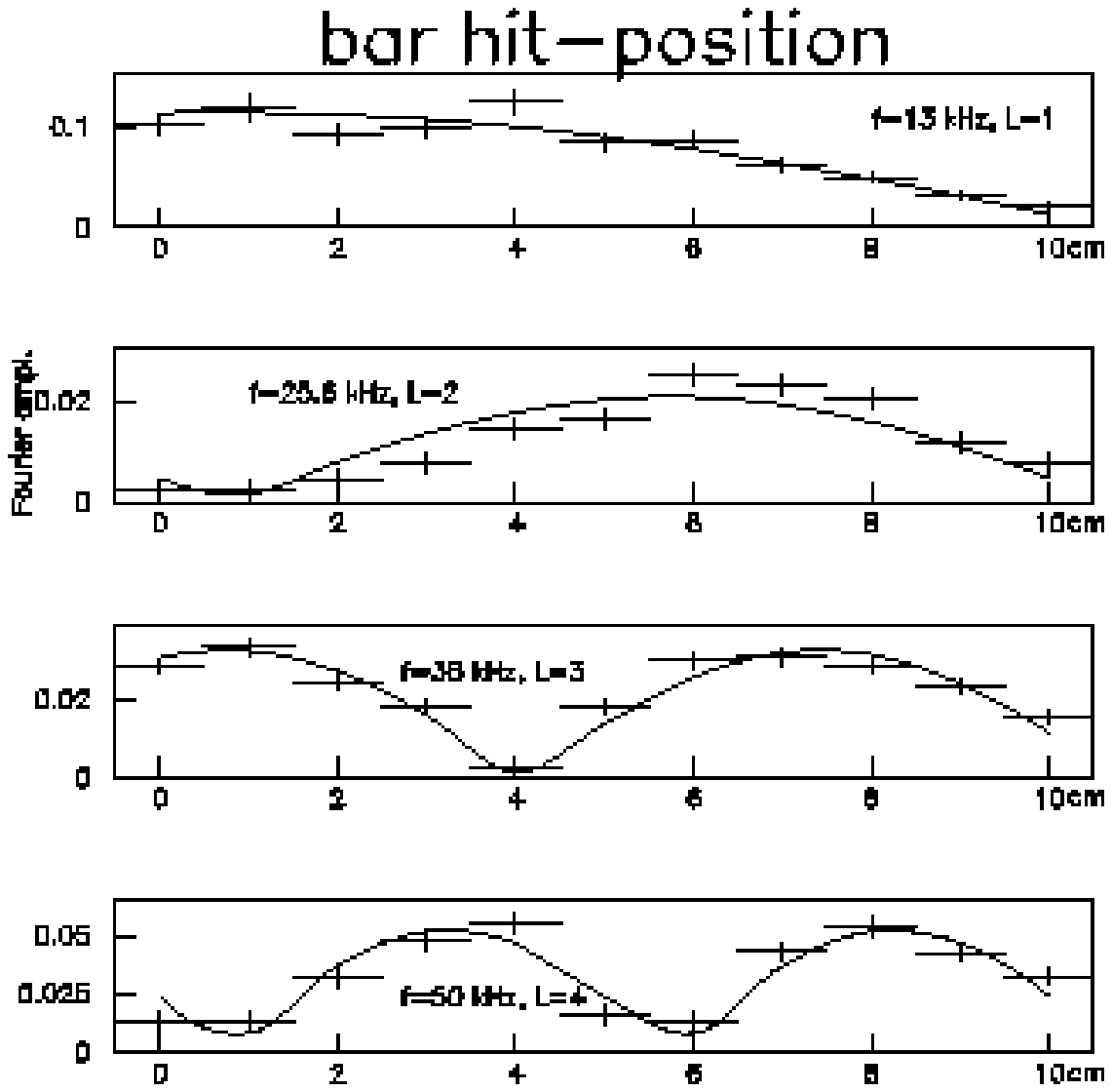,width=\fs cm}
\caption{
The measured, unnormalised
Fourier amplitudes ($+$) and model calculations (--)
as a function of the beam hit position along the cylinder axis
for the four lowest longitudinal modes of bar BC.
}
\label{fig:barhitpos}
\end{minipage}
\end{figure}
The Fourier amplitudes $A_k$ of $f(t)=\sum A_k e^{i\omega_k t}$ of the modes
depend linearly (as shown for the 13~kHz, $L$=1 mode in fig.~7)
on the integrated charge in
the beam pulse for a fixed beam position, and therefore also linearly on the
energy deposited by the beam, which ranged in these runs from 0.06-0.8~J.
The spread in the 
ratios of the amplitudes
to the beam charge, 
shows the Fourier amplitudes to reproduce
within $\pm~10\%$.

The agreement of the model to within 10\% with the measured
data is shown in 
fig.~8. 
The figure shows the measured Fourier amplitudes of bar BC
at the piezo sensor 
and the calculations following
Grassi Strini et al.~\cite{grassi,bernard} 
as a function of the hit position along the cylinder axis
for the first four longitudinal modes. For each mode the average
model value was scaled to the average measured value. 
The best fit was found with a
shift of the hit positions along the bar,
by an overall offset of $x_0$=-0.0075~m, which corresponds to the crude
way we aligned the bar with the beam line. 

\subsubsection{Lowest bar mode excitation amplitude}
\label{sub:lowbar}
\vspace{1.cm}
\Large
For the 13~kHz,~$L$=1 mode we determine the absolute amplitude for a comparison
with the model calculation of ref.~\cite{grassi,bernard}.
Firstly, we use the amplitude function B$_0$(x), see eq. 9 of
ref.~\cite{grassi,bernard},
by rewriting it in the form: 
\begin{eqnarray}
B_0(x)=2\cdot \kappa_0\cdot \Delta E/\pi 
\times cos(\pi x/l)sin(\pi\eta /(2l))/\pi\eta /(2l)
\label{eq:bzero}
\end{eqnarray}
with
\begin{eqnarray}
 \kappa_0=\alpha\cdot l/(c_v\cdot M)=\alpha/(c_v\cdot\rho O).
\label{eq:kap0}
\end{eqnarray}
In this expressions
$x$ is the hit position along the the cylinder axis, $l$ the bar length,
$\eta$ the beam diameter, $\alpha$ the thermal linear expansion coefficient, 
$\rho$ the density, c$_v$ the specific heat, 
$O$ the cylindrical surface area of the bar, and $\Delta E$ the energy
absorbed by the bar.
From $B_0(x)$ we derive the functional
form for the measured values of $W_{sens}$ as 
\begin{eqnarray}
W_{sens}(x)=\frac{B_0(x)}{\Delta E} \beta D\frac{dE}{dQ}
\end{eqnarray}
where $dE/dQ$ is the beam energy absorbed
by the bar per unit of impinging beam charge, $\beta$ the calibration factor
as discussed in section \ref{sec:calibration}, and D the decay 
factor $e^{-t/\tau}$, since eq. \ref{eq:bzero} applies at excitation time
and we have to correct the amplitude at measuring time 
for the mode's decay, corresponding to its Q-factor.  
Therefore,
\begin{equation}
 W_{sens}(x)=\kappa_{exp} 2/\pi
   \times cos(\pi x/l)sin(\pi\eta /(2l))/\pi\eta /(2l), 
\label{eq:wsen2}
\end{equation}
with
\begin{equation}
  \kappa_{exp}=\beta D\kappa_0 \frac{dE}{dQ}
\end{equation}
From fitting eq. \ref{eq:wsen2} to the measured values $W_{sens}(x)$ 
given in table II with $\kappa_{exp}$ as the free variable,
we find our presently measured value for
$\kappa_0^{exp}=\kappa_{exp}\cdot /(dE/dQ \cdot D \cdot \beta)$ 
which we compare to the model value in eq. \ref{eq:kap0}.
Secondly,
the decay time was measured
by recording the sensor signals after a trigger delayed by up to 
1.6~s at a
fixed beam hit position.
An exponential fit $A(t)=A_0*e^{-t/\tau}$ to the mode amplitude 
gives $\tau=(0.36 \pm 0.01)~$s for the $L$=1 mode. This
corresponds to a Q-value of $\approx$~15000, a value
consistent with the room temperature measurement of
aluminium as in ref.~\cite{Qval}, and indicating a negligible influence of the
suspension and piezoelectric ceramic sensor for this mode. 
From the measured value of
$\tau$ and a mean delay time
from the start of the beam pulse of 0.016~s, we calculate
the decay factor to be $D=0.95$.

\vspace{0.5cm}
\begin{minipage}{17 cm}
\small
{\Large
Table II. Excitation values W$_{sens}$, equalling the ratio of the
measured Fourier amplitude and the measured beam pulse charge at each of
the indicated hit positions on the bar for the 13 kHz, $L$=1 mode. 
}

\vspace{-0.2 cm}
\begin{center}
\begin{tabular}{|l|c|c|c|c|c|c|c|c|c|c|c|}
\hline
 hit position $x$ cm &0 &1 &2 &3 &4 &5 &6 &7 &8 &9 &10 \\
\hline 
 W$_{sens}$ V/nC &0.185 &0.216 &0.167 &0.180 &0.225 &0.152 &0.152 &
 0.157&0.112 &0.089 &0.057 \\
\hline
\end{tabular}
\newline
\end{center}
\end{minipage}
\vspace{0.5 cm}

Thirdly, as indicated, we use the data for W$_{sens}$
in the second row of table II to fit
the variable $\kappa_{exp}$ in eq. \ref{eq:wsen2}, 
where now x is the hit position as given in row~1,
$l$=0.2~m , and $\eta=0.002$~m.
The value found in the fit is 
$\kappa_ {exp}=(0.300\pm 0.025)$ V/nC.
Fourthly, from a Monte Carlo simulation 
at the beam energy of 570 MeV used for these runs, 
we calculate the mean absorbed energy and the mean energy spread
which results from the fluctuating energy losses of the passing electrons and 
the energies of the secondaries escaping from the bar,
as $\Delta $E$_e$=(19$\pm$~2) MeV.
The electron beam pulse thus deposits 
$dE/dQ = (0.019 \pm ~0.002)$ J/nC in the bar.
Using the measured calibration
value at f=12986 Hz as given in eq.~\ref{eq:ijk},
$\beta $=(2.2 $\pm $ 0.3) V/nm,
we arrive at
\begin{eqnarray}
  \kappa^{exp}_{0}= (7.4 \pm~ 1.4)~ \mbox{nm/J}.
\label{eq:kapbar}
\end{eqnarray}
Finally, we calculate the model value of $\kappa_0$
from the material constants 
as being $\kappa_0$=10~nm/J,
neglecting the much smaller error as arising from some uncertainty
in the parameters. We conclude that
$\kappa ^{exp}_0/\kappa_0= (0.74 \pm 0.14)$,
a result that is consistent with the validity of the model of ref.
\cite{grassi,bernard}. 
\newline
The measured maximum excitation amplitude at beam position $x$=0, see
fig.~8 
for the 13~kHz,~$L$=1 longitudinal mode thus corresponds 
to (0.13~$\pm$~0.02)~nm.
\newline

\subsubsection{Higher bar mode excitation amplitudes}
\label{sub:highbar}
Having determined the correspondence between the model calculation and the
experiment's result 
for the first longitudinal vibrational mode amplitude, we return to some of
the higher vibrational modes. 
To compare the modes we need to take the sensor position on the bar into
account. 
We rewrite the displacement amplitude of eq.~5 from ref. \cite{grassi}
as a function of hit position $x_h$ and sensor position $x_s$ as:
\[\Phi_{odd-L}=(2\kappa /L\pi)sin(L\pi x_s/l)cos(L\pi x_h/l), \]
\be
\Phi_{even-L}=(2\kappa /L\pi)cos(L\pi x_s/l)sin(L\pi x_h/l),
\label{eq:bcamps}
\ee
where $l$ is the bar length. We dropped the beam width correction term
which would lead to a less than 0.1\% correction even for $L$=4.
We approximate the sensor response by the local
strain along bar BC's cylinder axis, that is to the d$\Phi$/dx$_s$ of eq.
\ref{eq:bcamps}, arriving at a sensor response, S$_L$:
\begin{eqnarray}
 S_{odd-L}=B_L cos(L\pi x_h/l), 
 B_{odd-L}=(2\epsilon \kappa /l)cos(L\pi x_s/l), \nonumber \\
 S_{even-L}=B_L sin(L\pi x_h/l), 
 B_{even-L}=(2\epsilon \kappa /l)sin(L\pi x_s/l), 
\label{eq:bcsensresp}
\eea
where $\epsilon$ is a sensor response parameter.
The $x_s$ dependent term did
not enter into the calculation of
$\kappa_0^{exp}$ in the previous section,
since the calibration was done at the same sensor
position as the beam measurement. However,
for a comparison between the modes, the
dependence on the sensor position $x_s$ has to be taken into account.
Since the variables are strongly correlated, 
we, first, fitted for each mode the term $B_L$ in
the $x_h$ dependent part
of eq. \ref{eq:bcsensresp} to the measured value of $W_{sens}$ for the mode,
shifting the origin of $x_h$ by 0.0075~m, as mentioned before.
The results are given in the first row of table III.

\vspace{0.5 cm}
\begin{minipage}{17 cm}
\small
{\Large 
Table III.
Bar BC modes comparison. The piezoelectric ceramic sensor responds to the
bar's strain.  
}

\begin{center}
\begin{tabular}{|l|c|c|c|c|c|}
\hline 
 description & symbol & 13 kHz,$L$=1 & 25.6 kHz,$L$=2 & 38 kHz,$L$=3 & 50
 kHz,$L$=4 \\
\hline
 amplitude & $B_L^{meas}$ & 0.12$\pm 0.01$&0.021$\pm 0.002$&
  0.033$\pm 0.03$& 0.052$\pm 0.005$  \\
 decay correction & D &1.04$\pm $0.001&1.17$\pm 0.02$&
 1.49 $\pm $0.06&1.14$\pm $0.01\\  
 $B_L^{meas} \times D $ & $B_L^{exp}$ & 0.12$\pm 0.01$ & 0.025$\pm 0.003$
  &0.049$\pm 0.005$ & 0.059$\pm 0.006$\\   
 sensor position factor & $P_L$ & 1.02 & 2.61 & 1.20 & 1.41 \\ 
 $\epsilon \kappa 2/l =B_L^{exp}\times P_L (arb.u.)$ & $\kappa'$ &  
  0.12$\pm 0.01$ & 0.065$\pm 0.007$ & 0.059$\pm 0.006$ & 0.083$\pm 0.008$\\
\hline 
\end{tabular}
\end{center}
\label{tab:BCabs}
\end{minipage}

\vspace{0.5 cm}
Second, 
we corrected the amplitudes $B_L^{meas}$ for the mode decay with a factor $D$,
given in row~2, and corresponding to the times 
$\tau_1=0.36~s, \tau_2=0.10~s,~\tau_3=0.04~s,~\tau_4=0.12~s$, which leads to
the values of $B_L^{exp}$ in row~3.
Finally, we multiplied with the factor $P_{odd-L}=1/cos(L\pi x_s/l)$, 
$P_{even-L}=1/sin(L\pi x_s/l)$, where the bar length is $l$=0.2~m. 
Since the sensor extends from 0.005 through 0.020~m from the center of the bar,
we use the mean sensor position $x_s=0.0125$~m.
The resulting values of $\kappa'=2\epsilon\kappa /l$, shown in the last row,
should be independent of L. For $L$=2,3,4 they
are rather closely scattered
around a mean value of $\kappa'=0.07$ which is, however, at
about half the $L$=1 value.
This discrepancy might have originated
from some resonances of the sensor itself,
and we suspect the strong
peak at 23~kHz,
shown in figure~6, to be an indication of such
resonances playing a role.
\newline
\Large
Since the amplitudes of the higher modes for bar BC do not comply with
our expectations we turn, as a further check,
to our un-calibrated measurements with bar BU. 
It had been
equipped with a piezoelectric sensor 
at one end face where the longitudinal
modes have maximum amplitude. The sensor had been mounted   
flatly with about half of its surface glued to the bar, and responding to the
bar's surface acceleration, not its strain as at bar BC.
We extract the $\kappa _{L}$ values from our measurement
analogously as for bar BC,
following again the model calculations of
Grassi Strini et al.~\cite{grassi}, using the $L$=1 mode as the reference.
The results are given in table IV.

\vspace{0.5cm}
\begin{minipage}{17 cm}
\small
{\Large Table IV.
Bar BU modes comparison. The piezoelectric ceramic sensor responds to the
bar's acceleration.
The value of $\kappa_1^{meas}$ for the $L$=1, 13 kHz mode
is used as the reference for the higher modes.
}
\begin{center}
\begin{tabular}{|l|c|c|c|c|c|}
\hline 
 description & symbol & 13 kHz,$L$=1 & 25.6 kHz,$L$=2 & 38 kHz,$L$=3 & 50
 kHz,$L$=4 \\
\hline
 relative amplitude & $B^{meas}$ & 1 &  1.15 & 11.5 &  3.6  \\
 relative decay correction & D &1&3$\pm$1&0.7$\pm$0.2&1.3$\pm$0.9\\  
 $(\omega _{L=1}/\omega _L )^2$ & $\Omega$ & 1 & 0.26 & 0.12 & 0.07 \\
$B^{meas} \times D \times \Omega$ &
 $\kappa ^{meas}_L/\kappa^{meas}_1$ & 
   ${\bf 1}$&0.8$\pm$0.3&0.9$\pm$0.4&0.3$\pm$0.3\\   
\hline
\end{tabular}
\end{center}
\end{minipage}
\newline

\vspace{0.5cm}
\Large
After applying the decay correction factor $D$ and the frequency
normalisation factor $\Omega$, 
the results should be independent of L.
The $L$=4 value is significantly low,
which, again, might be due to some interfering resonance.
The $L$=2 and $L$=3 values, however, do not significantly deviate from
the $L$=1 value, thus confirming the model calculations for these higher
modes too.
\newline

\noindent
\subsection{Results for the sphere}
\label{sub:sphere}

\noindent
Our measurements on the sphere consisted of a) hitting the sphere with the beam
at one of two
heights in the vertically oriented plane through its suspension: at the equator
(E) and at 0.022~m southward (A);
b) rotating the sphere with its two fixed sensors
over $180^0$ around the suspension axis at each beam height,
and measuring several times back and forth 
by steps of $30^0$ to
diminish the influence of temperature and beam fluctuations, 
ending up on a $10^0$ angular lattice. 
\begin{figure}[h]
\vspace{-0.6cm}
\begin{minipage}[l]{\fs cm}
\psfig{figure=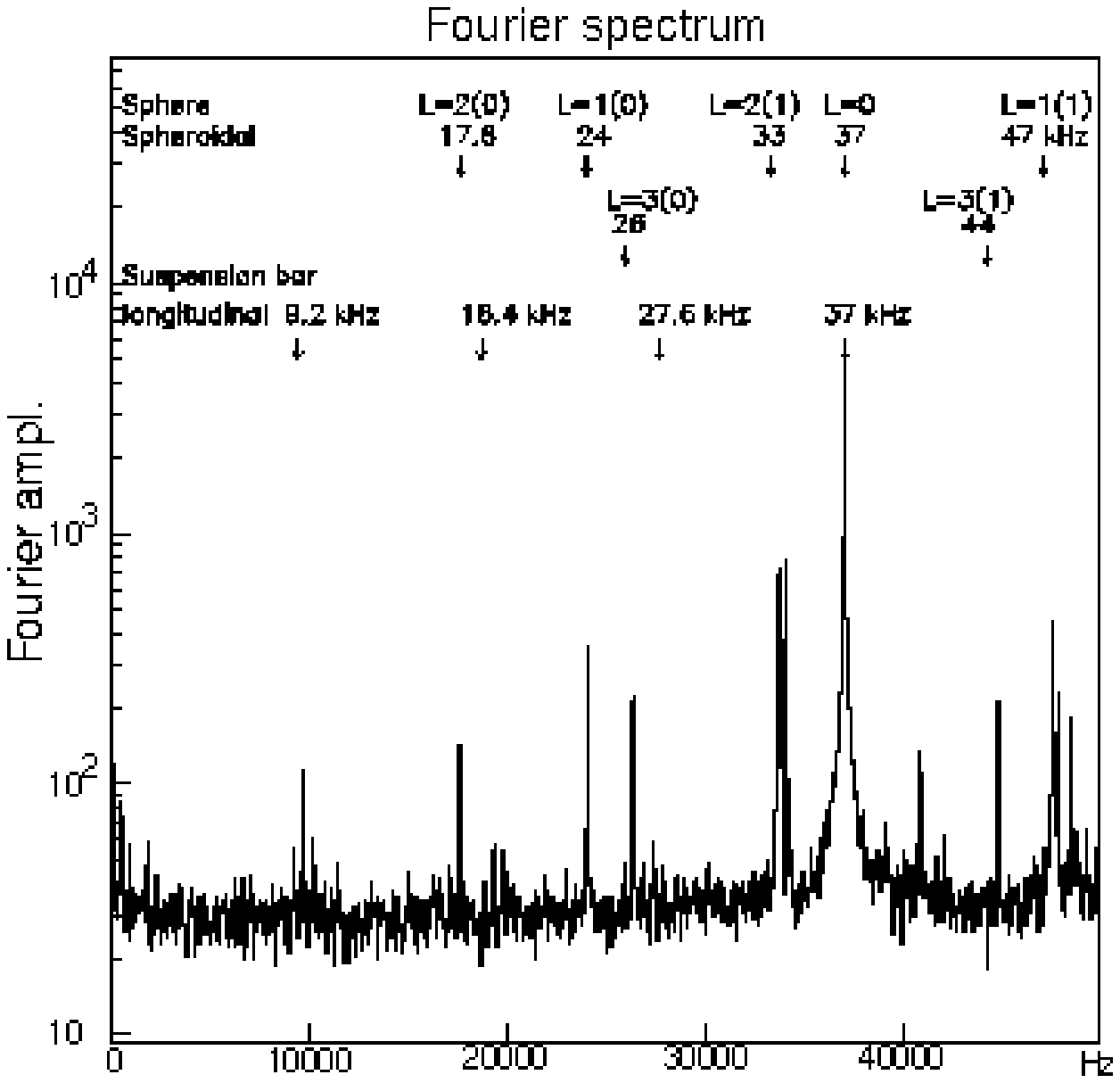,width=\fs cm}
\caption{
The Fourier amplitude spectrum of the sphere
SU averaged over all measured angles for sensor-1,
at beam height position E.
The modes and frequencies as calculated, are indicated.
}
\label{fig:bolFourier}
\end{minipage}
\begin{minipage}[l]{\fs cm}
\vspace{0.5cm}
\psfig{figure=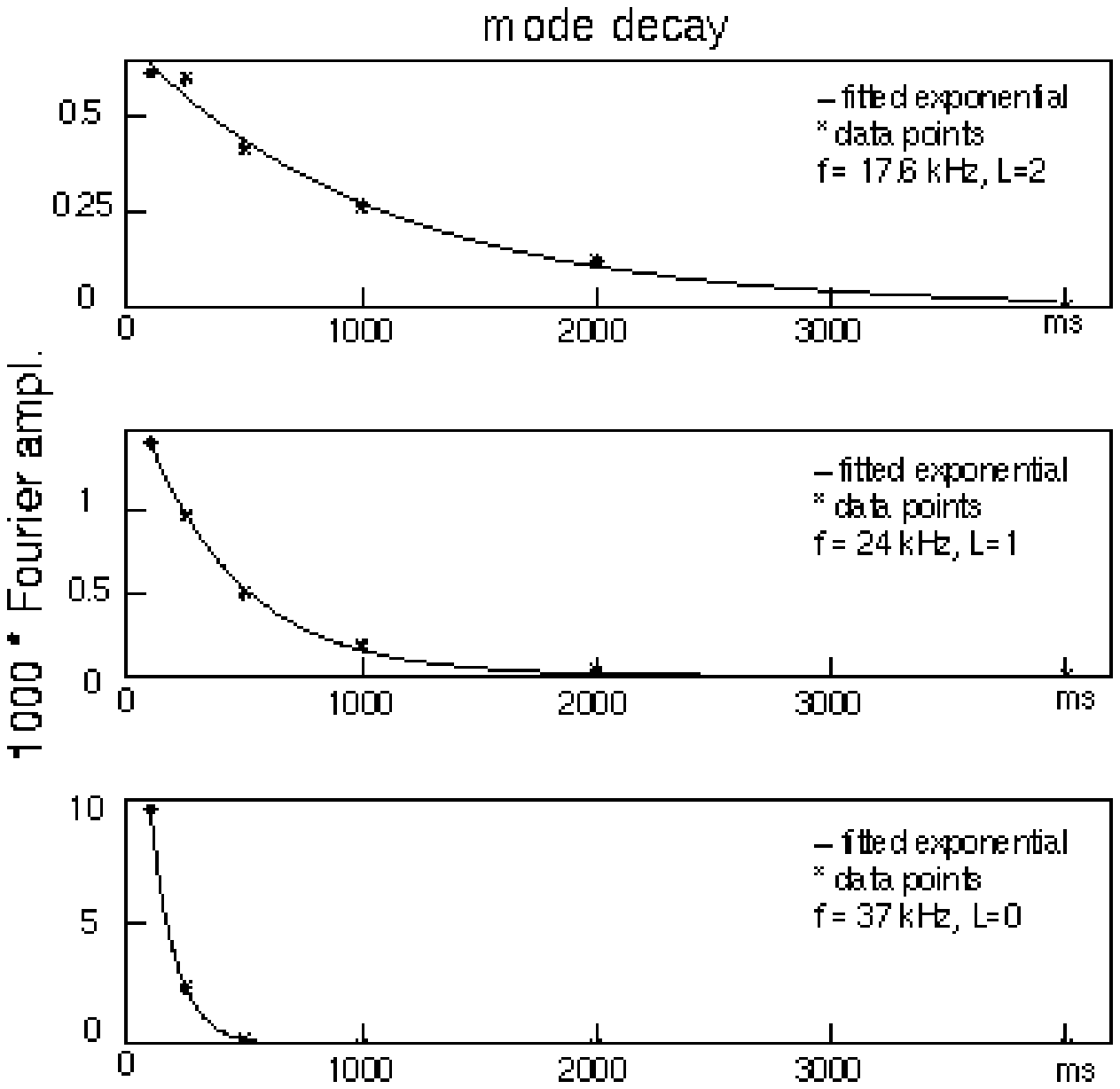,width=\fs cm}
\caption{
Decay of sphere's vibrational modes as measured with delayed data
taking of the spectrum analyser. The curve shows the fit of
$A(t)=A_0*e^{-t/\tau}$. 
Upper: 17.6 kHz, $L$=2, $\tau$=1.1 s.
Middle: 24 kHz, $L$=1, $\tau$=0.4 s.
Lower: 37 kHz, $L$=0, $\tau$=0.1s. 
}
\label{fig:sph-decay-17}
\end{minipage}
\end{figure}
The Fourier amplitude
spectrum of sensor-1, averaged over the angular positions,
is shown in fig.~9.
The lowest spheroidal mode is most relevant for a {\em spherical} resonant mass
gravitational wave detector, and we therefore focus on a few spheroidal modes.
As expected,
the lowest spheroidal $L$=2 mode is seen at 17.6~kHz,
the lowest spheroidal $L$=1 mode at 24~kHz, and
the lowest spheroidal $L$=0 mode at 37~kHz.
Some other peaks
are also indicated
in the figure, though not the toroidal modes, which we neglect completely.
It should be noted that
while the $L\neq $0 amplitudes oscillate over the angles, the $L$=0
amplitude does
not, leading to a relative enhancement of the latter in the angle-averaged 
fig.~9.
\newline
The Fourier amplitudes, again,
showed a linear dependence on the deposited energy.
\newline
To determine the decay times
at $f$=17.6, 24~kHz and 37~kHz, see fig.~10,
we took data with up to 4~s delay in the spectrum analyser, 
and found $\tau\approx 1~$s, 0.4~s and 0.1~s respectively.
\newline
\begin{figure}[h]
\vspace{-0.4cm}
\begin{minipage}[l]{\fs cm}
\psfig{figure=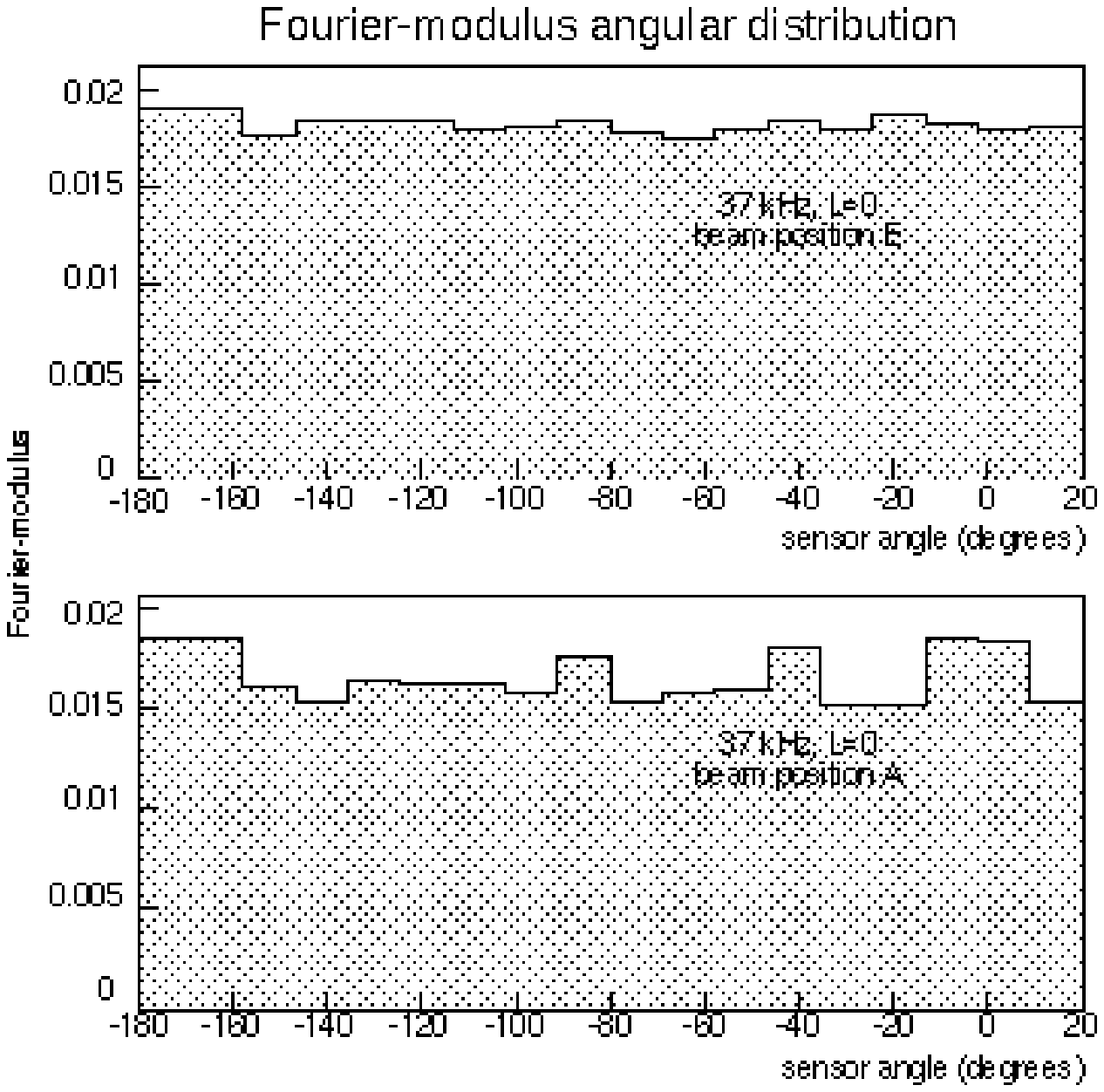,width=\fs cm}
\vspace{-0.5cm}
\label{fig:37hoekvsp}
\vspace{1.0cm}
\caption{
37~kHz Fourier-modulus angular distribution.
The beam hits the sphere at upper: E, lower: A. 
}
\end{minipage}
\hfill
\end{figure}
The angular distributions for the amplitude 
of the 37~kHz, $L$=0 mode at the two vertical
beam positions E and A are shown in figures~11.
The $L$=0 amplitude is independent of the angle, and
since the amplitude is constant to within 20\%
we infer
from the Fourier-modulus' deviation from flatness a
20\% variation of the beam intensity
from one shot to another. 
\newline
During our measurement with the sphere
we were unable to use the beam pulse as a trigger,
implying that the start time of data acquisition with respect to the beam
pulse is unknown. In our further analysis we will therefore use only
the Fourier modulus, and not analyse the phases.
\newline
The absolute scale of the 37~kHz Fourier amplitude turned out to be
$\approx 5$~times larger than the model value for sensor-2 and
$\approx 50$~times for
sensor-1. We assume this
discrepancy to be based on 
some interference effects, possibly with a sensor
resonance and a suspension bar mode, and we
do not further analyse the $L$=0 mode.
To disentangle the angular distributions in general,
we felt, would squeeze the results of our simple measurement too much,
for a couple of reasons.
First, the Fourier amplitude~$A_L$ for any multi-pole order $L$
in the sphere's case is actually a sum of
$M$-submodes. Though they would be degenerate for an ideal sphere, in practice
some $M$-modes might or might not turn out to be
split beyond the frequency-resolution of 
$\Delta f$=30~Hz. 
Second, 
both sensors $s_1,s_2$ should be taken to have
unknown sensitivities, {\bf e$_{s_j}$},
in three orthogonal directions, with phase factors +1 or -1 for their 
orientation.
Third, though each mode would start to be excited
within the same sub-nanosecond time interval
of the beam crossing, the building up of each mode's
resonance vibration may lead to a specific phase $t^0_{M_L,b_k}$
depending on the mode's spatial relation 
to the beam path. 
\newline
We now show that the calculated angular
distributions have the signature of 
the $L$-character of the measurement.
Therefore, we write
the Fourier modulus at different impinging beam positions $b_k$
as a function of the angle $\phi$ as: 
\begin{equation}
A_{L,s_j,b_k}(\phi)=\mid  F_{L,s_j,b_k}~
\mbox{\boldmath $\epsilon_{s_j}$} \cdot \sum_{-M_L}^{+M_L} s_{L,M_L,b_k} 
{\bf u}_{L,M_L,s_j,b_k}(\phi) e^{\omega_L t^0_{M_L,b_k}}\mid ,
\label{eq:mode}
\end{equation}
where $F_{L,s_j,b_k}$ is
a frequency response function for
each sensor, which may depend also on the beam position. This normalisation
factor is
expected to be of order 1, and is kept fixed at 1 for the $L$=2 distributions.
It is used as a free parameter for the $L$=1 distributions 
to compensate for the rather inaccurate knowledge of a) the sensor positions
on the sphere's surface, 
b) the beam track location and c) the 
electrons and photons shower development along the track, since
the exact excitation strengths of the modes
are quite sensitive to such data.
As the first step in the fitting procedure 
we separately calculated the $s_{L,M_L,b_k}{\bf u}_{L,M_L,s_j,b_k}$, where
$s_{L,M_L,b_k}$ is the mode's strength from
the beam excitation, as detailed in the appendix. 
We inserted the calculated $s_{L,M_L,b_k}{\bf u}_{L,M_L,s_j,b_k}$
in a hierarchical
fitting model, to simultaneously
fit~\cite{ref:minuit} the relevant parameters of eq. \ref{eq:mode} to
the 17.6 kHz, $L$=2 Fourier modulus $A_{L,s_j,b_k}$
for both sensors $s_1$ and
$s_2$ at both beam positions E and A. 
This fit led to a reduced $\chi^2$=1.3 at 59 degrees of freedom.
Next, with fixed values for the sensor efficiencies 
\mbox{\boldmath $\epsilon_{s_j}$} so
established, we fitted the relevant parameters for the
24 kHz, $L$=1 Fourier peaks, including the $L$=1 sensor response factors $F$. 
At all stages the
$\mid t^0_{M_L,b_k}\mid $ of the phases were kept within the bounds
of the period of mode-$L$.
With an uncertainty in the beam charge and in the Fourier peak amplitudes 
of $\approx 20\%$ each, 
the error amounts to $\approx $30\%, and we took
a minimum absolute error of 2$\times 10^{-5}$
for sensor $s_1$ and 1$\times 10^{-5}$ for sensor $s_2$.
In total we have 152 data points, while
the total number of fitted parameters is 27,
including a relative normalising factor
for the mean beam current at beam position A with respect to the mean current
at beam position E.
We found for the total fit a reduced $\chi^2$=1.6 at 125 degrees of freedom.
The $L$=1 response
factors remain within 1.1 and 0.2.
\begin{figure}[h]
\vspace{-0.4cm}
\begin{minipage}[l]{\fs cm}
\psfig{figure=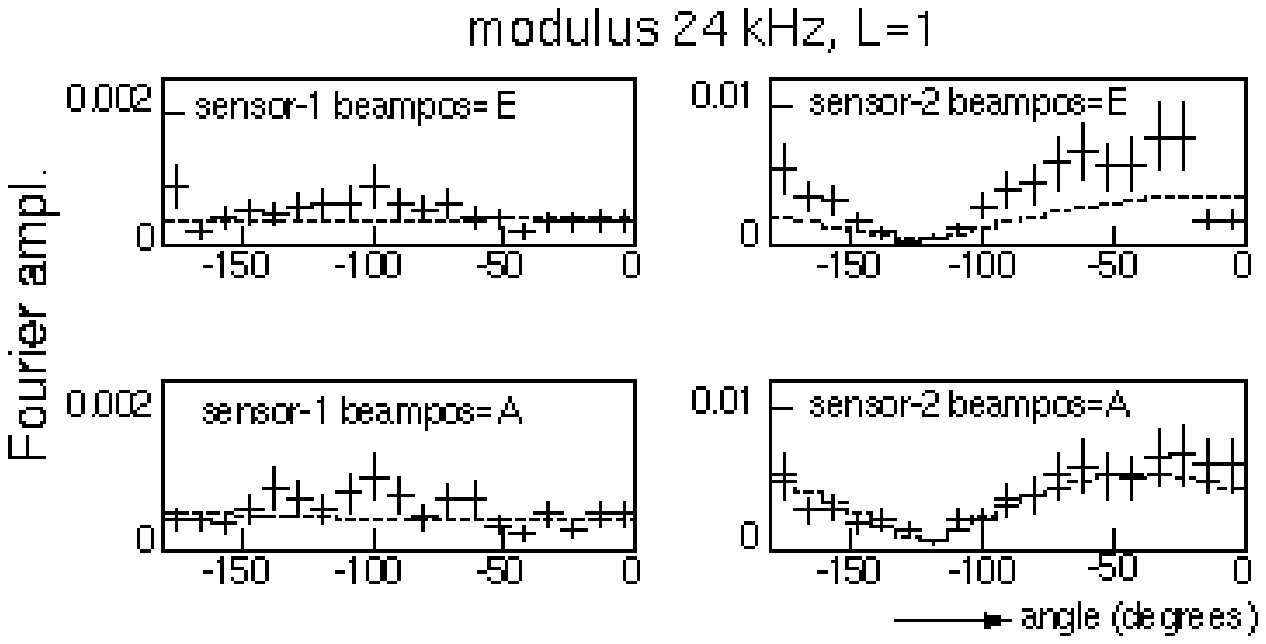,width=\fs cm}
\caption{
Data points (+) and
	fit results (---) for the sphere's 24 kHz, $L$=1 mode.  
	Left column: sensor-1, right colum: sensor-2.
	Upper row: beam position E, lower rows: beam position A. 
	The x-axes give the angle of sensor-1.
	Note that the y-scales are different for the two sensors.
}
\label{fig:fitsphere1}
\end{minipage}
\begin{minipage}[r]{\fs cm}
\psfig{figure=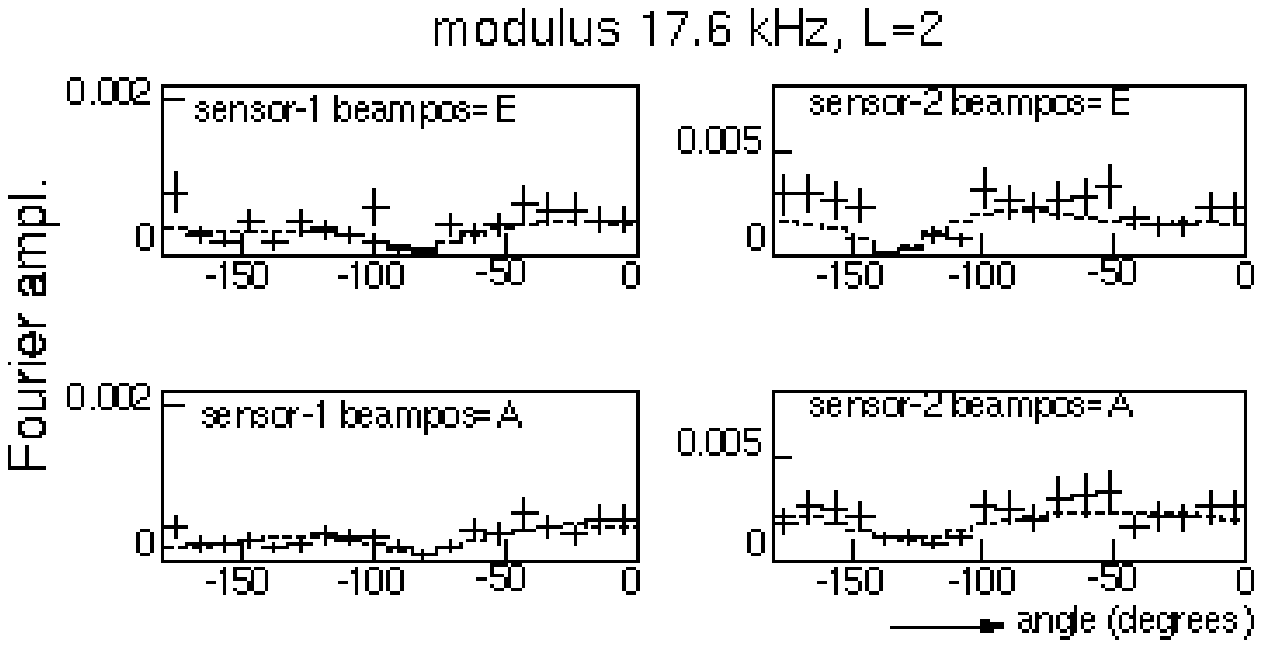,width=\fs cm}
\caption{
Data points (+) and fit results (---) for the sphere's 17.6 kHz, $L$=2 mode.  
	Left column: sensor-1, right colum: sensor-2.
	Upper row: beam position E, lower rows: beam position A. 
	The x-axes give the angle of sensor-1.
	Note that the y-scales are different for the two sensors.
}
\label{fig:fitsphere2}
\end{minipage}
\end{figure}
\Large
The results of the fits to the 17.6~kHz and 24~ kHz are given in figures
12 and 13
and table V. Note the different vertical scales 
used for sensor-1 and sensor-2 in both pictures. Some of the
parameters given in the table are strongly correlated.

\vspace{0.5cm}
\begin{minipage}[c]{17 cm}
\small
{\Large Table V. Results of a 
hierarchical fit to the data of the sphere at the 19 measuring angles of both
sensors $s_1$ and $s_2$ of the 
Fourier modulus at 17.6~kHz and 24~kHz at beam positions E and A.
}

\vspace{1 cm}
\begin{center}
\begin{tabular}{|l|c|c|c|c|c|}
\hline 
{\bf sensor efficiency} & fit result & error \\
&   $*10^{-4}$ V/ms$^{-2}$ & $*10^{-4}$ V/ms$^{-2}$ \\ 
\hline
$\epsilon_{s_1}-r$ & +0.10 & 0.02   \\
$\epsilon_{s_1}-\theta$ & -0.5 & 0.2   \\
$\epsilon_{s_1}-\phi$ & +0.4 & 0.1   \\
& & \\
$\epsilon_{s_2}-r$ & -0.5 & 0.2   \\
$\epsilon_{s_2}-\theta$ & -0.5 & 0.2   \\
$\epsilon_{s_2}-\phi$ & +4.0 & 0.3  \\
\hline
& & \\
$intensity(\frac{beam A}{beam E})$ & 0.7 & 0.1 \\ 
& & \\
\hline
{\bf response factor} &   & \\ 
\hline
$F_{L=1,s_1,beam A}$ & 0.3 & 0.1\\
$F_{L=1,s_2,beam A}$ & 1.1  & 0.1\\
& & \\
$F_{L=1,s_1,beam E}$ & 0.2  & 0.1\\
$F_{L=1,s_2,beam E}$ & 1.1  & 0.2\\
\hline
\end{tabular}
\end{center}
\end{minipage}

\vspace{0.5cm}
\Large
We conclude that
the measured Fourier amplitude angular distributions are consistent with
the model value for the $L$=2 and $L$=1 mode signature.
\newline

\subsubsection{The sphere's absolute displacement}
\label{subsub:absolute}
Finally, to estimate the order of magnitude
of the sphere's absolute displacement, we have to take an intermediate step
by first normalising bar BU to the calibrated results for bar BC and then use
bar BU as a calibration for the sphere. The
sensors used on bar BU consisting of the same sensor material and 
having been cut roughly to the same size, we assume to be identical
to the ones used on sphere SU. The
amplifiers used are identical. 
The bar BU sensors, however, differ strongly from those of bar BC.
\newline
We arrive at an indirectly calibrated
value for $\epsilon_{BU,2}= (3.4*10^{-4}\pm 30\%)$~V/ms$^{-2}$ 
of sensor-2 on
bar BU. The value of $\epsilon_{BU,1}$ for sensor-1 is about ten times smaller.
\newline
Then, for
the sphere, the fitted
value of \mbox{\boldmath $\epsilon_{s_j}$}
given in table V,
shows the largest value of sensor-2,
$\epsilon_{\phi-SU}=(4*10^{-4}\pm 10\%)~$V/ms$^{-2}$, to lead
to a ratio $\approx (1.2\pm 0.4)$ with 
$\epsilon_{BU}$. Again, the values for sensor-1 are about ten times
smaller. The error of $\approx 33\%$ is the propagated statistical error only.
The result seems reasonable.
So, the model calculation and our sphere measurement 
results are of the same order of magnitude on an absolute scale too.
\newline
From the maximum
Fourier modulus, $V^{max}=(0.003\pm 0.001)$~V, 
of the 17.6~kHz, $L$=2 sphere mode measured in
sensor sensor-2 as given in fig.~13, and the 
absorbed energy of 3.1~J,
we find the maximum sphere's
displacement to correspond to $(0.2\pm 0.1)$nm/J. 
\newline

\section{Discussion}
\label{sec:discussion}
Having confirmed the thermo-acoustic conversion model in the present
experiment, we discuss some points about
extrapolating these results to
the actual operation of a resonant gravitational wave detector. 
Firstly, in our experiment
many incident particles deposited their energy 
in the resonator, in contrast to a single 
muon hitting an actual detector. 
However, from this difference it seems unlikely to
reach different conclusions, especially since in the process of 
depositing energy along its track, the muon will generate lots of
secondary particles too.
Also, we
measured at room temperature while
actual detectors would have to operate in the millikelvin range. 
An aluminium resonator, for instance, at such a temperature,
would be superconducting, and it is as yet unclear how
the decoupling of the electron gas from the lattice would 
affect the process of acoustic excitation.
\newline
Therefore, we hold it of particular importance for
the prospected shielding of a next generation resonant mass gravitational wave
detector that an existing millikelvin detector like
the Nautilus~\cite{coccia,nautilus-eas}, 
would succeed in measuring
the impinging cosmic rays in correlation with the resonator mode.
Such a result, as a test for
the further applicability of the thermo-acoustic conversion model 
at operating temperature, would come
closest to the real situation envisaged for the new detectors.
\newline
Apart from such temperature effects,
the applicability of the thermal acoustic conversion model
\cite{beron,grassi,bernard}
is confirmed by the data and therefore
cosmic rays
should be expected to seriously disrupt,
as calculated by the model,
the possibility of detecting gravitational waves.
It is beyond the scope of this article to go into any detail \cite{exgrail}.
We want to 
point, however, to earlier 
calculations~\cite{mazzi-1,obers,ref:pizz} which, having used the model,
clearly show, firstly, that
a next generation spherical resonant mass gravitational wave detector
of ultra high sensitivity will be significantly excited by cosmic 
rays. Secondly,
the high impact rate of cosmic rays will prohibit gravitational
wave detection at the earth's surface, with the required sensitivity. 
\newline
Finally, shielding the instrument by 
an appreciable layer of rock as available in,
for instance, the Gran Sasso laboratory, 
would suppress the cosmic ray background by a factor of $\geq 10^6$.
Even then a vetoing system 
will be necessary and, at the radical reduction
of the background rate so established, it may indeed work effectively.

\section*{Acknowledgements}
We thank A.~Henneman 
for the computer code to calculate a sphere's vibrational modes, 
R.~Rumphorst for his knowledgeable estimate of sensor sensitivity,
J.~Boersma for digging out the formal orthogonality proof of a sphere's
eigenmodes, and the members of the former GRAIL team for expressing
their interest in this
study, especially P.W.~van Amersfoort,
J.~Flokstra, G.~Frossati, H.~Rogalla, A.T.M.~de Waele. 
This work is part of the research programme of 
the National Institute for Nuclear Physics
and High-Energy Physics (NIKHEF)
which is financially supported through
the Foundation for Fundamental
Research on Matter (FOM),
by the Dutch Organisation for Science Research (NWO).


\pagebreak
\section*{Appendix:
Sphere excitation model calculation}
Our calculation of the ($L,M$)-mode excitation strengths
is based on the source term of eq. 5.10/11 of ref.
\cite{bernard}, $s=\Sigma/(\rho V) *\int dz \nabla_{\perp}\cdot {\bf u}$,
with $\Sigma=\gamma dE/dz$. Here, $\gamma$ is the Grueneisen constant,
$\rho V$ de sphere's mass, and $dE/dz$ the absorbed energy per unit track
length.
The Fourier amplitudes, measuring the second time derivative of the mode
amplitudes, are directly proportional to $s$, and the mode amplitudes follow
from $s/\omega^2$, as in eq. 5.18 of \cite{bernard}.
However, the amount of energy absorbed per unit length in our 
case depends on the particle's position along the track.  
We therefore re-included the $\Sigma$-term under the source term's 
integral by letting $dE(z)/dz$ represent the 
electromagnetic 
cascade development of ref.~\cite{ref:bijbel}
as an approximation to the amount
of energy absorbed per unit track length
by the sphere at position z along the beam
track, 
\begin{equation}
s_{L,M_L}=\kappa \int_L {\bf \nabla}_{\bot}\cdot 
  {\bf u}_{L,M_L}(z)\frac{dE(z)}{dz} dz, 
\label{eq:source}
\end{equation}
where $z$ is being 
measured from the beam's entrance point into the sphere.
With 
$E_{abs}$ being the total amount of
energy absorbed by the sphere from the electron bunch,
we write $dE(z)/dz=E_{abs}*d(E(z)/E_{abs})/dz$ 
and use the polynomial expansion
$d(E(z)/E_{abs})/dz=\sum_{i=0}^3 c_i z^i$.
Then $\int_{z_{min}}^{z_{max}} dz d(E(z)/E_{abs})/dz$=1. 
For the polynomial, measuring $z$ in meter, we acquired the values
$c_0$=0.8332~m$^{-1}$, $c_1$=226~m$^{-2}$, $c_2$=-1832~m$^{-3}$,
$c_3$=4909~m$^{-4}$
from a fit to the form given in 
ref. \cite{ref:bijbel}, with less than a percent deviation for our case
of $0\leq z\leq 0.15$~m.
The value for the energy absorbed by the sphere from a single electron,
$E_{abs}^e$=123~MeV, we got 
from both our Monte Carlo simulation
using GEANT \cite{ref:geant}, and from
EGS4 \cite{ref:EGS4,ref:richw}. At the measured 25~nC beam pulse charge this
corresponds to a total $E_{abs}=3.1~$J absorbed by the sphere.
Then the value 
of $\kappa~E_{abs}= \gamma E_{abs}/M$=1.00~m$^2$/s$^2$, 
for our case of $M=\rho V$=4.95 kg 
and $\gamma$=1.6.
Our sphere has a suspension hole which leads to a slight shift
in the frequencies and the spatial distribution
of the modes, with respect to those of a sphere without a hole \cite{hole}. 
We approximated, however, our sphere's modes by 
the ideal hole-free sphere's eigenmode solutions {\bf u}($z$) \cite{orthog},
using the available computer code as established
in ref.~\cite{ref:alex}, and renormalising to 
$\int {\bf u}\cdot {\bf u}dV=V$, as used
in ref.~\cite{bernard} from eq. 5.6 onward.
The source term $s_{L,M_L}$
was calculated for each mode ($L,M$) by numerically integrating
eq. \ref{eq:source}. We checked the surface term in the numerical procedure
to be negligible, as assumed
in the partial integration leading to the form of $s$ used by
ref.~\cite{bernard}.
Each {\bf u}($\phi$) in eq. \ref{eq:mode} is the eigenmode solution,
calculated for each sensor 
on the $\phi$-grid of the measured data,
and each term $s_{L,M_L,b_k}$ is the 
excitation factor $s_{L,M_L}$ at the specific beam position $b_k$.


\begin{thebibliography}{99}
\bibitem{grail}
P.~Astone, G.V.~Pallottino, M.~Bassan, E.~Coccia, Y.~Minenkov, I.~Modena,
A.~Moleti, M.A.~Papa, G.~Pizzella, P.~Bonifazi, R.~Terenzi, M.~Visco,
P.~Carelli, V.~Fafone, A.~Marini, S.M.~Merkowitz, G.~Modestino, F.~Ronga,
M.~Spinetti, L.~Votano,
in: 
p. 551 in: E.~Coccia, G.~Pizzella, G.~Veneziano (eds.),  
Proc. 2nd Amaldi conf. on
Gravitational Waves, CERN 1997, World Scientific 1999.
\\C.~Frajuca, N.S.~Magalhães, O.D.~Aguiar, N.D.~Solomonson, W.W.~Johnson,
S.M.~Merkowitz and W.O.~Hamilton, 
in: Proc. OMNI-I(1996), World Scientific 1997,
\\ G.M. Harry, T.R. Stevenson, H.J. Paik, Phys.Rev.D54(1996)2409
\\ C. Zhou, P.F. Michelson, Phys.Rev.D51(1995)2517.
\\ GRAIL, NIKHEF, Amsterdam, May 1997.
\\ G.D.~van~Albada, W.~van~Amersfoort, H.~Boer~Rookhuizen, J.~Flokstra,
G.~Frossati, H.~van~der~Graaf, A.~Heijboer, E.~van~den~Heuvel,
J.W.~van~Holten, G.J.~Nooren, J.E.J.~Oberski, H.~Rogalla,
A.~de~Waele, P.K.A.~de~Witt~Huberts,
     in Proc. Second Workshop on Gravitational Wave Data
     Analysis, pag. 27, Editions Fronti\`eres, 1997.
\bibitem{ricci} F. Ricci, Nucl.Instr. and Meth. A260(1987)491,
\\J. Chang, P. Michelson, J. Price, Nucl.Instr. and Meth. A311(1992)603
\bibitem{mazzi-1}
G. Mazzitelli, M.A. Papa, Proc. OMNI-I(1996), World Scientific 1997
\bibitem{beron}
B.L. Beron, R. Hofstadter, Phys.Rev.Lett. 23,4(1969) 184, \\
B.L.~Beron, S.P.~Boughn, W.O.~Hamilton, R.~Hofstadter, T.W.~Martin,
IEEE Trans.Nucl.Sc. NS17(1970)65.
\bibitem{grassi}
A.M. Grassi Strini, G. Strini, G. Tagliaferri, J.Appl.Phys. 51,2(1980)948, 
\bibitem{askar}
For non-resonant thermo-acoustic effects see for instance
I.A. Borshkoysky, V.V. Petrenko, V.D. Volovik, L.L. Goldin, Ya.L. Kleibock,
M.F. Lomanov, Lett.Nuovo Cimento 12(1975)638 \\
L. Sulak, T. Armstrong, H. Baranger, M. Bregman, M. Levi, D. Mael, J. Strait,
T. Bowen. A.E. Pifer, P.A. Polakos, H. Bradner, A. Parvulescu, W.V. Jones, 
J. Learned, Nucl.Instr. and Meth. 161(1979)203 \\
J.G. Learned, Phys.Rev.D 19,11(1979)3293 \\
G.A. Askariyan, B.A. Dolgoshein, A.N. Kalinovsky, N.V. Mokhov,
Nucl.Instr. and Meth. 164(1979)267. 
\bibitem{obers} 
J.E.J. Oberski, J.W. van Holten, G. Mazzitelli, 
Assessing the effects of cosmic rays on a resonant-mass gravitational
wave detector, NIKHEF-97/3
\bibitem{ref:amaldi}
G.D.~van~Albada, 
H.~van~der~Graaf, G.~Heijboer, 
J.W.~van~Holten, W.J.~Kasdorp, J.B.~van~der~Laan,
L.~Lapik\'{a}s, G.J.L.~Nooren, C.W.J.~Noteboom, J.E.J.~Oberski, 
H.Z.~Peek, A.~Schimmel, T.G.B.W.~Sluijk, J.~Venema, P.K.A.~de~Witt~Huberts,
p. 402 in: E.~Coccia, G.~Pizzella, G.~Veneziano (eds.),  
Proc. 2nd Amaldi conf. on
Gravitational Waves, CERN 1997, World Scientific 1999.
\bibitem{strooi}
C. de Vries, C.W.~de~Jager, L.~Lapik\'{a}s, G.~Luijckx, R.~Maas,
H.~de~Vries, P.K.A.~de~Witt~Huberts,
Nucl.Instr. and Meth. 223(1984)1
\bibitem{meaweg}
In the beginning of 1999 the Amsterdam MEA electron accelerator, and AmPS
stretcher ring ended
their operations permanently, due to stopped funding. Several  
parts are being dispersed over labs in Europe, Russia and the USA.
\bibitem{DvA-bar}
J.F. de Ronde, G.D. van Albada and P.M.A. Sloot
in High Performance Computing and Networking '97, 
     Lecture Notes in Computer Science, pag. 200, Springer, 1997. \\ 
J.F. de Ronde, G.D. van Albada and P.M.A. Sloot
Computers in Physics, 11(5):484--497, Sept/Oct 1997.\\ 
J.F. de Ronde. Mapping in High Performance Computing,
     PhD. thesis Univ. of Amsterdam, The Netherlands, 1997.\\ 
J. de Rue. On the normal modes of freely vibrating elastic objects of
     various shapes, thesis, Univ. of Amsterdam, the Netherlands, 1996. 
\bibitem{freq-long}
H. Kolsky, Stress waves in solids, Dover 1961, \\
A.E.H. Love, Mathematical theory of elasticity, Dover 1944, \\
D. Bancroft, Phys.Rev. 59(1941)588,\\
R.Q. Gram, D.H.~Douglas, J.A.~Tyson, Rev.Sci.Instr. 44,7(1973)857
\bibitem{metal-hb}
Metals Handbook, vol. 6,
9th edition 1983, American Society of Metals, Metal park, OH.
\bibitem{bernard}
D. Bernard, A. de Rujula, B. Lautrup, Nucl.Phys. B242(1984)93.
\bibitem{Qval}
W. Duffy, J.Appl.Phys. 68(1990)5601.
\bibitem{ref:minuit}
MINUIT fitting tool, CERN, http://wwwinfo.cern.ch/asd/cernlib/minuit
\bibitem{coccia}
E. Coccia, A.~Marini, G.~Mazzitelli, G.~Modestino, F.~Ricci, F.~Ronga,
L.~Votano,
Nucl.Instr. and Meth. A355(1995)624
\bibitem{nautilus-eas}
P. Astone, M.~Bassan, P.~Bonifazi, P.~Carelli, E.~Coccia, V.~Fafone,
A.~Marini, G.~Mazzitelli, S.M.~Merkowitz, Y.~Minenkov, I.~Modena,
G.~Modestino, A.~Moleti, G.V.~Pallottino, M.A.~Papa, G.~Pizzella,
F.~Ronga, M.~Spinetti, R.~Terenzi, M.~Visco, L.~Votano,
LNF-95/35, Nucl.Phys.B(Proc.Suppl.)70(1999)461.
\bibitem{exgrail}
The Dutch subsidising agencies NWO/FOM have
decided against further pursuing the GRAIL project.
They favoured more conventional, on-going
work, above
the funding of our team's proposal
to research, develop and open up a new field in The Netherlands
with the GRAIL resonant sphere Gravitational Wave detector, even
though an evaluation
committee of international experts gave GRAIL an
almost embarrassingly positive judgement.
We therefore see, sadly,
no opportunity for a follow up to the current paper.
\bibitem{ref:pizz}
G.Pizzella, "Do cosmic rays perturb the operation of a large resonant
spherical detector of gravitational waves?", LNF-99/001(R)
\bibitem{ref:bijbel}
D.E. Groom, Passage of particles through matter, in \\ 
Rev.Part.Phys~C3,1-4(1998)148
\bibitem{ref:geant}
GEANT~simulation~tool,~CERN, http://wwwinfo.cern.ch/asd/geant/index.html
\bibitem{ref:EGS4}
Electron Gamma Shower development, OMEGA project, \\
http://www-madrad.radiology.wisc.edu/omega/www/omega\_intro\_00.html
\bibitem{ref:richw}
We thank Richard Wigmans for running the case. 
\bibitem{hole}
At an even smaller frequency than the
first quadrupole mode of the full sphere,
another quadrupole mode arises when the sphere has a spherical
hole, actually being a thick spherical shell.
We did not consider
the latter in our study, however, since it has its maximum amplitude at the
inner surface and a minimum amplitude at the outer surface.
\bibitem{orthog}
M.E. Gurtin, The Linear Theory of Elasticity, sec. E.VI., The free vibration
problem, p. 261 in: C. Truesdell (ed.), Mechanics of Solids, Handbuch der
Physik VIa/2, Springer 1972, 
proves for an ideal sphere, even if partially clamped, 
the orthogonality of its modes.
\bibitem{ref:alex}
A.A. Henneman, J.W. van Holten, J. Oberski, 
Excitations of a wave-dominated spherical resonant-mass detector,
NIKHEF-96-006
\end{thebibliography}
\end{document}